\documentclass[twocolumn,showpacs,preprintnumbers,amssymb,prl]{revtex4}        
\usepackage{graphicx}
\usepackage{color}
\usepackage{dcolumn}
\usepackage{bm}
\usepackage{longtable}
\date{\today}
\begin{document}
\title{Swapping path-spin intraparticle entanglement onto spin-spin interparticle entanglement}

\author{S. Adhikari}
\altaffiliation{satyabrata@bose.res.in}
\affiliation{S. N. Bose National Centre for Basic Sciences,
Salt Lake, Kolkata 700 098, India}

\author{A. S. Majumdar}
\altaffiliation{archan@bose.res.in}
\affiliation{S. N. Bose National Centre for Basic Sciences,
Salt Lake, Kolkata 700 098, India}

\author{Dipankar Home}
\altaffiliation{dhome@bosemain.boseinst.ac.in}
\affiliation{CAPSS, Dept. of Physics, Bose Institute, Salt Lake, 
Kolkata-700091, India}

\author{A. K. Pan}
\altaffiliation{apan@bosemain.boseinst.ac.in}
\affiliation{CAPSS, Dept. of Physics, Bose Institute, Salt Lake, 
Kolkata-700091, India}

\begin{abstract}
Based on a scheme that produces an entanglement between the spin and the path variables of a single spin-$1/2$ particle (qubit) using a beam-splitter and a spin-flipper, we formulate a procedure for transferring this intraparticle hybrid entanglement to an interparticle entanglement between the spin variables of two other spatially separated spin-$1/2$ particles which never interact with each other during the entire process. This procedure of entanglement swapping is accomplished by a Mach-Zehnder setup in conjunction with the Stern-Gerlach measuring device and by using suitable unitary operations. The proposed protocol, thus, enables the use of intraparticle entanglement as a resource - a feature that has remained unexplored.
\end{abstract}


\maketitle

Over the years, quantum entanglement has been increasingly recognised as a key ingredient in
the information theoretic processes involving storage and distribution of information among the
fundamental constituents of the world \cite{peres1}. The first
profound implication of entanglement for quantum foundations was noticed
way back in 1935 using position and momentum  variables
\cite{EPR}, and was later extended to the discrete spin variables
\cite{bohm}. In recent times there has been significant development
of the theory of entanglement for systems described by various types
of Hilbert spaces, such as those corresponding to discrete variables\cite{review1} as well as for those pertaining to continuous variables
\cite{review2}. Several powerful information processing protocols
such as cryptography \cite{crypto}, dense-coding \cite{denscod},
quantum teleportation \cite{bennett1}, and entanglement swapping \cite{swap} have been developed for the spin entangled states,
as well as for the position-momentum entangled states \cite{teleport,cvcrypt}.

The study of various aspects of entanglement and the associated applications in the context of information theory promises to
provide insights into a wide range of diverse phenomena such as phase
transitions in condensed matter systems \cite{condmat} and black hole
physics \cite{bhinf}. Against this backdrop, the investigation of \textit{hybrid entanglement} between
the physical variables in mutually different Hilbert spaces
such as those corresponding to the path (or linear momentum) variables on the one hand, and spin variables on the other, is
of special relevance. Although the theoretical framework of quantum 
mechanics allows for the existence of hybrid entangled states involving
Hilbert spaces with distinct properties, the possibility of
physical realization of such states has not been much explored and is only 
beginning to be appreciated \cite{hybrid}.

Another interesting recent line of development is based on the idea of generation of intraparticle entanglement
between the different degrees of freedom of the {\it same particle}. The 
entanglement between polarization and linear momentum of a single photon \cite{polmom}, and also that between polarization and angular
momentum of a single photon \cite{polangmom} have been demonstrated 
experimentally. The idea of creating an entanglement between the path and the
spin degrees of freedom for a single spin-$1/2$ particle was first proposed
in order to demonstrate a testable incompatibility between quantum mechanics and noncontextual realist models \cite{home}. Subsequently, such a path-spin hybrid entangled state for a single neutron has been realized experimentally \cite{hasegawa}.  Recently,  an interesting application of hybrid intraparticle entanglement has been discussed in the context of neutrino oscillations \cite{dell}. There have also been studies that use the notion of the intraparticle entanglement for demonstrating nonlocality of a single photon \cite{van}.

Now, an important point to note is that since intraparticle entanglement between the different degrees of freedom 
is confined to a single particle, such an entanglement should be relatively easier to preserve,
at least in principle, against decoherence effects. It is
then natural to ask the question whether this type of hybrid entanglement
between the different degrees of freedom of the \textit{same} particle can be used as a
\textit{resource} for information processing. 

At the outset, the above idea may seem difficult to implement,  since the entanglement considered is \textit{not} shared nonlocally 
between two spatially separated regions in a way that is amenable to be exploited as a resource. 
One  way out, however, would be to transfer a given intraparticle entanglement onto an entanglement between the appropriate degrees of freedom of
two spatially separated particles. 

It is from this perspective that in this paper we demonstrate \textit{how} the path-spin entanglement of a single spin-$1/2$
particle can be transferred to the spin-spin entanglement involving two  spin-$1/2$ particles which remain spatially separated and non-interacting with each other during the entire procedure by which this transfer of entanglement is achieved. In order to outline the realizability of such a scheme, we first discuss a method to generate an intraparticle hybrid path-spin entanglement using a beam-splitter and a spin-flipper. Then, our protocol for entanglement swapping becomes implementable with the help of two additional spin-$1/2$ particles that never interact with each other during the entire process. For this, two separate parties (Alice and Bob) perform a series of operations that include unitary transformations together with appropriate measurements involving the Stern-Gerlach devices, followed by the use of classical communication that enables to eventually transfer the information content of an intraparticle path-spin entangled state to the spin-spin entanglement pertaining to two spatially separated spin-$1/2$ particles.

The proposed scheme is pictorially illustrated in Figure 1. 
\vskip -1.0cm
\begin{figure}[h]
%{\rotatebox{0}{\resizebox{9.0cm}{7.5cm}{\includegraphics{fig.eps}}}}
{\rotatebox{0}{\resizebox{9.0cm}{7.5cm}{\includegraphics{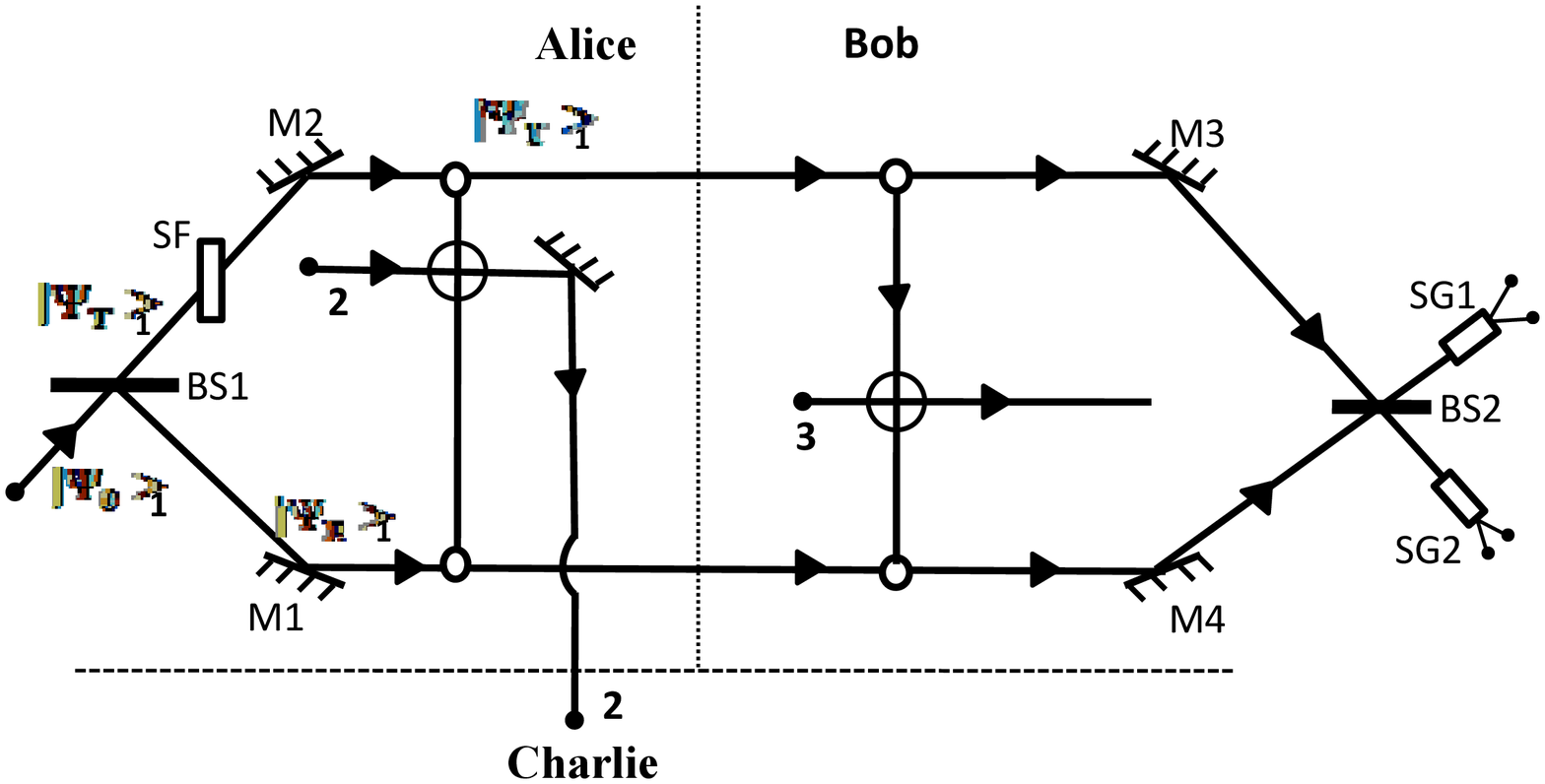}}}}
\end{figure}
\vskip -2.0cm
{\footnotesize FIG. 1: A spin-$1/2$ particle (labelled as particle 1) falls on the beam-splitter BS1. A spin-flipper is placed along the transmitted channel. A CNOT operation is performed by Alice involving particle `2' and  the particle `1' which is subsequently transported to Bob. While the particle `2' is sent to Charlie, Bob recombines the reflected and the transmitted channels corresponding to particle 1 by using the beam-splitter BS2, and then Bob performs spin measurements using the Stern-Gerlach devices SG1 and SG2. According to the measurement results, Bob performs appropriate unitary transformations on the states of the two qubits `1' and `3' that are with him. As a result, the two spatially separated qubits `2' and `3' with Charlie and Bob respectively get entangled although they have never interacted with each other.} \\

Let us consider a spin-$1/2$ particle (say, a neutron, labelled as particle 1) 
that has an initial spin polarized state along the $+\widehat {z}-axis$ (denoted 
by $\left|\uparrow_{z}\right\rangle$). Taking into consideration its path (or
position) variables, the joint path-spin state can be written as 
\begin{eqnarray}
|\Psi_{0}>_{ps} = |\psi_{0}\rangle_{p} \otimes |\uparrow_{z}\rangle_{s}
\label{redstate}
\end{eqnarray}
where the subscripts $p$ and $s$ refer to the path and the spin variables respectively.
The ensemble of neutrons corresponding to $|\Psi_{o}>_{ps}$ are with Alice, incident on a  beam-splitter (BS1) whose 
transmission and reflection probabilities are $|\alpha|^{2}$ and $|\beta|^{2}$ 
respectively. Any given incident particle can then emerge along either the 
reflected  or the transmitted channel corresponding to the state designated 
by $\left|\psi_{R}\right\rangle$ or $\left|\psi_{T}\right\rangle$ respectively.

Here we recall that for any given lossless beam-splitter, arguments using the 
unitarity condition show that for the particles incident on the beam-splitter, 
the phase shift between the transmitted and the reflected states of particles 
is essentially $\pi/2$\cite{zeilinger}. For simplifying our treatment, we 
take the transmission and reflection amplitudes of BS1 and BS2 to be 
real quantities where $\alpha^{2}+\beta^{2}=1$. Note that the beam-splitter acts only on the path-states without
affecting the spin-state of the particles, while in our argument a crucial role is played by the mutually 
orthogonal path states $\left|\psi_{R}\right\rangle$ and 
$\left|\psi_{T}\right\rangle$ which are  eigenstates of the projection 
operators $P(\psi_{R})$ and $P(\psi_{T})$ respectively. 

These projection operators  can be regarded as corresponding to the observables that pertain 
to the determination of \textit{`which channel'} a particle is found to be in. 
For example, the results of such  a measurement for the reflected (transmitted)
channel with binary alternatives are given by the eigenvalues of 
$P(\psi_{R})$ ($P(\psi_{T})$); the eigenvalue $+1 (0)$ corresponds to a 
neutron being found (not found) in the channel represented by 
$\left|\psi_{R}\right\rangle$ ($\left|\psi_{T}\right\rangle$).

The state of a particle emergent from BS1 can then be written as
\begin{eqnarray}
|\Psi_{0}>_{ps}\rightarrow |\Psi_{1}>_{ps} =
(\alpha|\psi_{T}\rangle_{p}+i\beta|\psi_{R}\rangle_{p}) \otimes
|\uparrow_{z}\rangle_{s} \label{transformedstate}
\end{eqnarray}
The state vectors
$|\psi_{T}\rangle_{p}$, $|\psi_{R}\rangle_{p}$, $|\uparrow_{z}\rangle_{s}$,
and $|\downarrow_{z}\rangle_{s}$ can be represented as
\begin{eqnarray}
|\psi_{T}\rangle_{p}\equiv\left(\begin{matrix}{0 \cr 1
}\end{matrix}\right), \>\>
|\psi_{R}\rangle_{p}\equiv\left(\begin{matrix}{1 \cr 0
}\end{matrix}\right),\nonumber\\
|\uparrow_{z}\rangle_{s}\equiv|0\rangle_{s}=\left(\begin{matrix}{0
\cr
1}\end{matrix}\right), \>\>
|\downarrow_{z}\rangle_{s}\equiv|1\rangle_{s}=\left(\begin{matrix}{1
\cr 0 }\end{matrix}\right)\label{notation}
\end{eqnarray}
At this stage, we may stress that although here a single particle is considered at a time, the dichotomic
path and spin variables enable it to be viewed effectively as two qubits.
Using the above notation, the state (\ref{transformedstate}) can be 
written as
\begin{eqnarray}
|\Psi_{1}>_{ps} = \left(\begin{matrix}{0 \cr i\beta \cr 0 \cr \alpha
}\end{matrix}\right)\label{vector}
\end{eqnarray}
Next, let us suppose that the particles in the channel corresponding to 
$|\psi_{T}\rangle$ pass through a spin-flipper (SF) (that contains a uniform magnetic field along, 
say, the $+\widehat x$-axis) which flips the state $\left|\uparrow\right\rangle_{z}$ to $\left|\downarrow\right\rangle_{z}$. 
As a consequence of this insertion of a spin-flipper
in one of the channels, the particle 1 with Alice has now the path-spin entangled state given by
\begin{eqnarray}
|\Psi>_{ps} &=& \alpha|\uparrow_{z}\rangle_{s}\otimes|\psi_{T}\rangle_{p}
+i\beta|\downarrow_{z}\rangle_{s}\otimes|\psi_{R}\rangle_{p}\nonumber\\
&\equiv&
\alpha|0\rangle_{s}\otimes|\psi_{T}\rangle_{p}+i\beta|1\rangle_{s}\otimes|\psi_{R}\rangle_{p}
\label{pathspinent}
\end{eqnarray}
Note that the above path-spin entanglement, given by Eq. [\ref{pathspinent}], is between the spin variables and 
the pseudo-spin like path observables of a spin-$1/2$ particle -  this is what we call the hybrid `intraparticle  
entanglement' that is distinct from the usually discussed 
`interparticle entanglement', say,  between the spin variables of two 
spatially separated particles, as well as is different from the recently discussed form of
`hybrid entanglement' \cite{hybrid} between the polarization of one photon and the
spin of another spatially separated photon.
 
Next, we assume that Alice possesses \textit{another particle} (labelled as particle 2)
which is in the up-spin state $|\uparrow\rangle_{2}\equiv|0\rangle_{2}$.
The total state is now the combination of  the path-spin entangled state 
and the spin state $|0\rangle_{2}$, which can be written  as
\begin{eqnarray}
|\Omega\rangle_{ps2} &=& |\Psi\rangle_{ps}\otimes|0\rangle_{2} \nonumber\\
&=& \alpha|00\rangle_{s2}\otimes|\psi_{T}\rangle_{p}
+i\beta|10\rangle_{s2}\otimes|\psi_{R}\rangle_{p}
\label{composite}
\end{eqnarray}
At this stage, we consider that Alice performs a suitable two qubit CNOT operation on the qubits 's' and '2' by taking the qubit `s' as a source and the qubit `2' as a target qubit.
Then, as a result, the transformed state is given by
\begin{eqnarray}
|\Phi\rangle_{ps2}
=\alpha|00\rangle_{s2}\otimes|\psi_{T}\rangle_{p}
+i\beta|11\rangle_{s2}\otimes|\psi_{R}\rangle_{p} \label{cnota}
\end{eqnarray}

Next, let us suppose that  Alice  sends the particle `1' (embodying the qubits `p' and `s') to her distant
partner Bob who possesses another particle '3' with the spin polarised state $|\uparrow_{z}\rangle$. After
receiving the particle '1', Bob performs a CNOT operation on the qubits 's'
and '3' on his side by taking the qubit 's' as a source and the qubit '3' as
a target qubit. Taking into account Bob's operation, the four-qubit 
joint (Alice-Bob) state can be written as
\begin{eqnarray}
|\Psi\rangle_{ps23}=
\alpha|000\rangle_{s23}\otimes|\psi_{T}\rangle_{p}
+i\beta|111\rangle_{s23}\otimes|\psi_{R}\rangle_{p} \label{cnotb}
\end{eqnarray}
where note that the qubits `p', `s' and `3' are physically
with Bob whereas the qubit `2' with Alice is sent to another distant party Charlie.

Subsequently,  the particle 1 passing through either of the two channels 
$\left|\psi_{T}\right\rangle$  ($\left|\psi_{R}\right\rangle$) is reflected 
by the mirrors M2 and M3 ( M1 and M4) - these reflections do not lead to any 
net relative phase shift between the channels $\left|\psi_{R}\right\rangle$ and
$\left|\psi_{T}\right\rangle$.
Bob then uses a $50-50$ beam splitter (BS2) to recombine the two paths. 
The states $|\psi_{T}\rangle_{p}$ and
$|\psi_{R}\rangle_{p}$ are transformed by BS2 to
\begin{eqnarray}
|\psi_{T}\rangle_{p}\rightarrow
\frac{1}{\sqrt{2}}(i|\psi_{T}^{\prime}\rangle_{p}+|\psi_{R}^{\prime}\rangle_{p}){}\nonumber\\
|\psi_{R}\rangle_{p}\rightarrow
\frac{1}{\sqrt{2}}(|\psi_{T}^{\prime}\rangle_{p}+i|\psi_{R}^{\prime}\rangle_{p})
\label{trans}
\end{eqnarray}
In writing Eq.(\ref{trans}) we have taken into account a relative phase 
shift of $\pi/2$ between the states $\left|\psi_{R}^{\prime}\right\rangle$ and 
$\left|\psi_{T}^{\prime}\right\rangle$ that  arises because of the reflection 
from BS2.
Using the transformation (\ref{trans}), the four-qubit state represented
by Eq.(\ref{cnotb})
evolves to
\begin{eqnarray}
|\Psi\rangle_{ps23} &=&
\frac{i}{\sqrt{2}}[(\alpha|000\rangle_{s23}+\beta|111\rangle_{s23})\otimes|\psi_{T}^{\prime}\rangle_{p}]\nonumber\\
&+& \frac{1}{\sqrt{2}}[(\alpha|000\rangle_{s23}-\beta|111\rangle_{s23})\otimes|\psi_{R}^{\prime}\rangle_{p}]
\label{bs}
\end{eqnarray}

Beyond the beam-splitter BS2, the Stern-Gerlach apparatuses denoted by SG1 and SG2
are placed {\bf by which the qubit `s' undergoes a unitary transformation}
$|0\rangle_s \rightarrow \frac{1}{\sqrt{2}}(|0\rangle_s + |1\rangle_s)$,
$|1\rangle_s \rightarrow \frac{1}{\sqrt{2}}(|0\rangle_s - |1\rangle_s)$. 
Now, depending upon which of the two paths $|\psi_{T}^{\prime}\rangle_{p}$ or
$|\psi_{R}^{\prime}\rangle_{p}$ are taken by Bob's particle 1, there
exist the following possibilities:\\

Case-I: If Bob's particle 1 travels along $|\psi_{T}^{\prime}\rangle_{p}$ then after its interaction with
the Stern-Gerlach apparatus, the reduced three-qubit state is given by
\begin{eqnarray}
|\Psi\rangle_{s23}=
\frac{1}{\sqrt{2}}[|0\rangle_{s}\otimes(\alpha|00\rangle_{23}+\beta|11\rangle_{23})\nonumber\\
+|1\rangle_{s}\otimes(\alpha|00\rangle_{23}-\beta|11\rangle_{23}]
\label{sg10}
\end{eqnarray}
In this case, the following are the possibilities:\\

(i) If Bob's measurement outcome is $|0\rangle_{s}$, then the two-qubit 
state for the paticles 2 and 3 is given by
\begin{eqnarray}
|\chi_{1}\rangle_{23}=
\alpha|00\rangle_{23}+\beta|11\rangle_{23}
\label{out1}
\end{eqnarray}
(ii) If the measurement outcome is $|1\rangle_{s}$, then the two-qubit 
state at for 2 and 3 reduces to
\begin{eqnarray}
|\chi_{2}\rangle_{23}=
\alpha|00\rangle_{23}-\beta|11\rangle_{23}
\label{out2}
\end{eqnarray}
Case-II: If Bob's particle 1 travels along
$|\psi_{R}^{\prime}\rangle_{p}$, then after its interaction with
the Stern-Gerlach apparatus, the reduced three-qubit state is given by
\begin{eqnarray}
|\Psi\rangle_{s23}=
\frac{1}{\sqrt{2}}[|0\rangle_{s}\otimes(\alpha|00\rangle_{23}-\beta|11\rangle_{23})\nonumber\\
+|1\rangle_{s}\otimes(\alpha|00\rangle_{23}+\beta|11\rangle_{23}]
\label{sg20}
\end{eqnarray}
In this case, the following are the possibilities:\\

(i) If Bob's measurement outcome is $|0\rangle_{s}$, then the two-qubit 
state for the particle 2 and 3 is given by
\begin{eqnarray}
|\chi_{3}\rangle_{23}=
\alpha|00\rangle_{23}-\beta|11\rangle_{23}
\label{out3}
\end{eqnarray}

(ii) If the measurement outcome is $|1\rangle_{s}$, then the two-qubit 
state for 2 and 3 reduces to 
\begin{eqnarray}
|\chi_{4}\rangle_{23}=
\alpha|00\rangle_{23}+\beta|11\rangle_{23}
\label{out4}
\end{eqnarray}

At this final stage, all that remains for Bob to do is to perform suitable unitary transformations to create an entangled state of the qubits `2' and `3' that are with Bob and Charlie respectively  - this state being a replica (in terms of the information content) of the 
path-spin entangled state (\ref{pathspinent}) originally possessed by Alice. The required unitary operations corresponding to each measurement outcome
are given in the following Table:\\

\textsl{Table-1: }\\\\
\begin{tabular}{| c| c| c| }
\hline
  Path state & Measurement   & Unitary transformation \\
             & outcome       & performed on qubits 2 and 3\\
   \hline
$|\psi_{T}^{\prime}\rangle_{p}$ & $|0\rangle_{s}$ & $I\otimes S$\\
 & $|1\rangle_{s}$ &  $I\otimes \sigma_{z}.S$ \\ \hline
$|\psi_{R}^{\prime}\rangle_{p}$ & $|0\rangle_{s}$ & $I\otimes \sigma_{z}. S$ \\
 & $|1\rangle_{s}$ & $I\otimes S$  \\ \hline
 \end{tabular}\\\\
Here $\sigma_z$ is the Pauli spin matrix and $S$ denotes the phase gate 
defined as
\begin{eqnarray}
S = \left(\begin{matrix}{1 & 0 \cr 0 & i}\end{matrix}\right)
\label{phasegate}
\end{eqnarray}
Thus, after performing the unitary operations as indicated in Table 1, for each measurement outcome, Bob and Charlie are ultimately left with sharing a two-qubit spin-spin interparticle entangled state which has
the same information content as that embodied in the original path-spin 
intraparticle entangled state used by Alice.

Therefore, the  upshot of our above demonstration is that the ``swapping'' of the intraparticle entanglement encoded in particle `1' into useful interparticle entanglement of the spatially separated particles `2' and `3' is achieved by satisfying the following key condition: Particles `2' and `3' are kept separated and they never interact with each other during the entire process. In other words, using our scheme, the particles `2' and `3' can get entangled by only letting the particle `1' to interact independently with the particles `2' and `3', without the need of having `2' and `3' in the same place and without the need for a joint unitary operation on `2' and `3'.

Before concluding, we may note that Cubitt \emph{et al.}\cite{cubitt} had suggested a scheme using which two distant particles can be entangled by continuous interaction with a mediating(ancilla) particle that never itself becomes entangled. It should, however, be evident that our scheme is formulated in a way that is  basically  different from that proposed by Cubitt \emph{et al.} because, in essence, what our work reveals is the procedure by which the information encoded in the entanglement between two different degrees of freedom of the \textit{same} particle can be
\emph{transferred} across a distance by creating an entangled spin state of two spatially separated particles. This procedure, therefore, opens up the possibility that the hybrid entanglement at the level of a single particle can be used to perform interesting information processing tasks. Such a possibility needs to be explored by further detailed studies. 

It is worth stressing the ubiquity of the path (or linear momentum) degrees of freedom in any experimental setup using particles/photons. In our proposed scheme, it is the pseudo-spin like path variable that has been invoked to first generate the path-spin entanglement for a single qubit. Then, using a suitable setup, this entangled state is transferred  to an entanglement between the spins of two qubits which remain spatially separated and never interact with each other. While two CNOT operations have been used in two different stages during the process, they are essentially local operations that ultimately enable nonlocal sharing of entanglement between two spatially separated particles without requiring them to be subjected to any global unitary operation. {\bf It may be noted here that
it has been proposed in \cite{gui} that the path information could be used as
a qubit source to simulate a duality quantum computer, thus providing a possible
application of our scheme for such purposes.} Further, note that, although the suggested protocol is demonstrated for spin-$1/2$ particles, it may  also be implemented using, say, photons and suitable optical devices. Finally, this demonstration underscores the usability of path-spin entanglement pertaining to a single particle \cite{home,hasegawa} as a physical resource, thereby exemplifying  the power of  hybrid entanglement \cite{hybrid} as a fundamental concept, independent of any particular physical realization of Hilbert space.

{\it Acknowledgements:} ASM and DH acknowledge support from the DST Project 
 SR/S2/PU-16/2007. DH also thanks the Centre for Science, Kolkata for support.

\end{document}